\documentclass[vecphys]{svmult}

\usepackage{makeidx}         % allows index generation
\usepackage{graphicx}        % standard LaTeX graphics tool
\usepackage{multicol}        % used for the two-column index
\usepackage[bottom]{footmisc}% places footnotes at page bottom
\usepackage[numbers,sort&compress]{natbib}
\makeindex             % used for the subject index
\begin{document}

\title*{Recent Results from HERA}
% Use \titlerunning{Short Title} for an abbreviated version of
% your contribution title if the original one is too long
\author{Andrea Parenti}
\institute{Universit\`a degli Studi di Padova and INFN Sezione di Padova,\\
via Marzolo 8, 35131 Padova, Italy.\\
\texttt{parenti@pd.infn.it}
}

%
% Use the package "url.sty" to avoid
% problems with special characters
% used in your e-mail or web address
%
\maketitle

\section{Introduction}
\label{sec:introduction}
 The HERA collider located in Hamburg, Germany, is a unique facility
 which collides protons of $920$~GeV
 \footnote{Until 1998, the energy of the proton beam was $820$~GeV.}
 and electrons (or positrons) of $27.5$~GeV.

 The $e^\pm p$ interactions proceed via a $\gamma$/$Z^0$
 exchange in the neutral current (NC) reaction, or via a
 $W^{\pm}$ exchange in the charged current (CC) interaction.
% The kinematic of the process is described by the $e^\pm p$ center of
% mass energy squared, $s$, and by two out of three Lorentz invariant
% quantities, $Q^2$: the absolute value of the invariant mass squared of
% the exchanged particle, $x$: the fraction of the proton momentum
% carried by the struck quark, and $y$: the fractional energy
% transferred to the proton in its rest frame. These variables are
% related through $Q^2=sxy$, if the masses of the electron and the
% proton are neglected.

 HERA is taking data since 1992; after a shutdown in 2000, data taking
 has restarted in 2002 with a five-fold increase in the luminosity.
 The first running period is referred to as HERA-I, whereas the
 second one is referred to as HERA-II.

 In this report I will review some recent results obtained by the H1 and ZEUS
 experiments at the HERA collider.
 Many different processes can be studied at HERA;
 this selection reflects my personal taste
 and many other relevant results have been omitted.

\section{Results}
\label{sec:results}
 {~\\ \bf Structure Functions and Parton Distribution Functions}

\begin{figure}
\centering
\includegraphics[width=6.3cm]{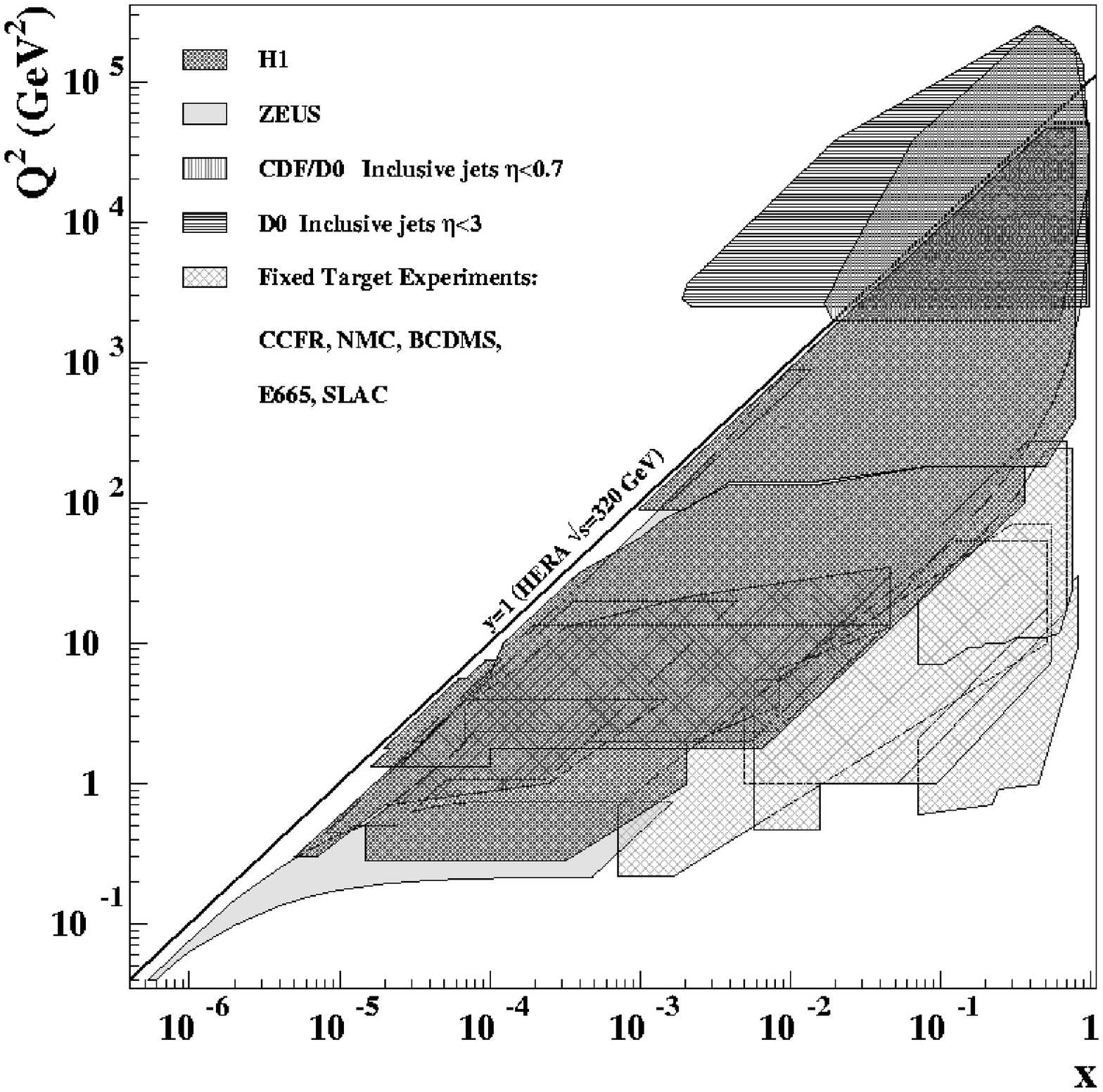}%
\includegraphics[width=6.3cm]{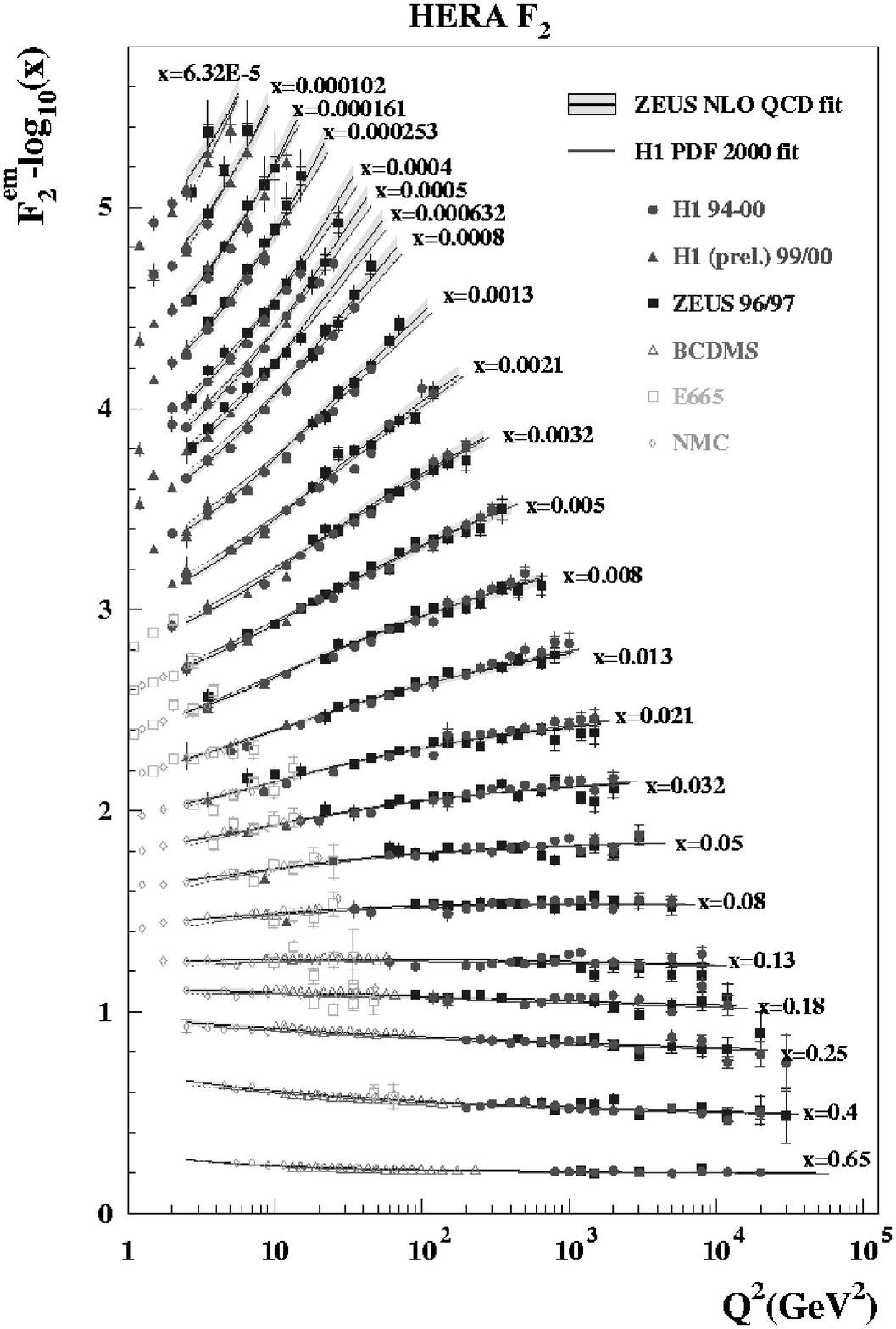}
\caption{The $x$-$Q^2$ kinematical plane and the accessible regions
 at hadronic colliders, in fixed target experiments and at HERA (left).
 The structure function $F_2$ versus $Q^2$ at fixed values of
 $x$, compared with a fit based on DGLAP equations (right).}
\label{fig:pdf} % Give a unique label
\end{figure}

 The cross section for the $ep \to eX$ process with unpolarised lepton beams
 depends on the proton structure functions $F_2$, $F_L$ and $F_3$;
 the prominent contribution comes from $F_2$.
 HERA is the ideal machine to study proton structure functions, since
 the accessible kinematical region is much larger than in hadronic colliders
 and fixed target experiments (see Fig.~\ref{fig:pdf}).

 $F_2$ was measured by using HERA-I data \cite{Adloff:2003uh,Chekanov:2001qu}
 in the region $6.32\times10^{-5}<x<0.65$ and $1<Q^2<30000$~GeV$^2$
 with an uncertainty as little as 2--3\% (see Fig.~\ref{fig:pdf}).
 The longitudinal structure function $F_L$ has been measured
 as well, though with worse precision \cite{hep-ph-0311180}.
 $F_3$, the parity violating part of the interaction,
 is measurable as a difference in $e^-p$ and $e^+p$
 NC differential cross sections.
 A measurement from ZEUS \cite{hep-ph-0607116}
 -- using $e^-p$ collisions from HERA-II and $e^+p$ from HERA-I --
 has been presented at DIS 2006;
 this measurement has a much lower statistical uncertainty
 than with HERA-I data only.

 The charm and beauty contributions to $F_2$ have been extracted
 from HERA-I data by H1 \cite{epj:c45:23,epj:c40:349} and
 ZEUS \cite{pr:d69:012004}.
 H1 used the impact parameter significance in order to evaluate
 the heavy quark fraction in the sample; in the region $Q^2>150$~GeV$^2$
 and $0.1<y<0.7$ the charm and beauty contributions to $F_2$ are
 $18.3\pm3.0~\%$ and $2.72\pm0.74~\%$, respectively.
 ZEUS at the time did not have any vertex detector and therefore
 identified only charm by the decay chain $D^* \to D^0 \pi_s$,
 $D^0 \to K \pi$. The charm contribution to $F_2$ goes up to $\sim 30~\%$
 at $Q^2=500$~GeV$^2$ and $x=0.012$.
 A vertex detector was installed in the ZEUS detector before HERA-II phase,
 therefore ZEUS will be able to evaluate $F_2^{b\bar{b}}$ with the new data.

 The proton parton distribution functions (PDFs) were extracted by
 ZEUS \cite{epj:c42:1} from the NC and CC cross sections,
 jet production in $e^+p$ collisions,
 dijet production in $\gamma p$ collisions, by using only HERA-I data
 and no other data from different experiments (the so-called ZEUS-JET fit).
 The results are already competitive to MRST and CTEQ fits.
 The importance of the ZEUS-JET fit is that the uncertainty is dominated
 by statistics and will be reduced significantly by the use of HERA-II data,
 whereas in MRST and CTEQ the uncertainty comes mainly from systematics.

% {~\\ \bf First Results With ZEUS Vertex Detector}
%
% The ZEUS vertex detector (MVD) was installed in 2001 inside the central
% tracker. The MVD consists of a barrel part, organized in three cylindrical
% layers, and a forward section, with four wheels perpedicular to the beam
% direction.
%
% The first analysis done by means of the MVD is the measurement of beauty
% photoproduction in 33 pb$^{-1}$ of $e^+p$ collisions collected in
% 2004 \cite{pos:hep2005:070}.
% A sample having two jets and a muon in the final state was selected, and the
% beauty content extracted by the simultaneous fit of the muon impact parameter
% and the $p_T^{rel}$ variable\footnote{$p_T^{rel}$ is the transverse momentum
% of the muon with respect to the jet axis.}.
% The $d \sigma / d p_T^{\mu}$ was evaluated and found compatible
% with the ZEUS HERA-I measurement and the FNMR prediction.

 {~\\ \bf First Results with Polarised Lepton Beam}

 The leptons at HERA acquire spontaneously a transverse polarisation
 due to the synchrotron light emission.
 In 2003 spin rotators were installed before H1 and ZEUS interaction
 points, transforming the transverse polarisation in longitudinal.
 The CC cross section depends linearly on the degree of polarisation
 of the beam, $P_e$:
 $\sigma_\mathrm{CC}^{e^\pm p} =
 (1\pm P_e)~\sigma_\mathrm{CC,unpol}^{e^\pm p}$.

 H1 and ZEUS have measured \cite{pl:b634:173,pl:b637:210}
 the total cross section of the CC process for $e^+p$ collisions
 and different values of polarisation and found a nice
 agreement with the expected $(1+P_e)$ dependence.

 The NC dependence on $P_e$ is more complicated;
 ZEUS has measured \cite{pl:b637:210} the $d \sigma /dQ^2$
 in $e^+p$ collisions with positive and negative values of the polarisation
 and found again a nice agreement with the Standard Model prediction.

 {~\\ \bf Isolated Leptons at High $P_T$}

 Events with missing $P_T$ and the production of an isolated, high-$P_T$ lepton
 come mainly from the process $ep \to eWX$, $W \to l \nu$ ($l=e,~\mu,~\tau$);
 investigations of the process with $l=e,~\mu$ have been performed by both
 the H1 and ZEUS experiments \cite{pos:hep2005:319}.
 H1 observed an excess of high-$P_T$ isolated leptons in HERA-I $e^+p$ data
 with respect to the expectation from $W$ production; this excess was not
 seen by ZEUS.
 The analysis of HERA-II data confirms the HERA-I results:
 H1 observed 15 events in 1994-2004 $e^+p$ data having hadronic $p_T > 25$~GeV
 ($4.6\pm 0.8$ expected), whereas no excess was seen in $e^-p$ collisions
 and by ZEUS (in both $e^\pm p$).
 These results are not yet understood.

 {~\\ \bf Measurement of $\alpha_S$ at HERA}

 The strong coupling constant, $\alpha_S$, appears in all QCD processes:
 $\alpha_S$ can therefore be extracted from many observables.

 A precise estimation of $\alpha_S$ at $M_Z$ scale has been
 made \cite{aipcp:792:689} by averaging the measurements made at HERA
 (e.g. jet cross section, event shape, jet multiplicity, etc.);
 the energy dependence of $\alpha_S$ was measured, too.
 The uncertainty of HERA determination is dominated by the theoretical part,
 whereas the experimental part is already better than the world average
 total uncertainty.

 {~\\ \bf Search for Pentaquarks at HERA}

 A number of experiments observed a narrow baryon resonance with positive
 strangeness, mass around 1530~MeV, decaying to $nK^+$ or $p (\bar{p}) K^0_S$.
 The signals are consistent with an exotic state having quark content
 $uudd \bar{s}$, called $\Theta^+$.
 Other experiments did not observe any signal.

 ZEUS observed a peak \cite{pl:b591:7} in the $p (\bar{p}) K^0_S$
 invariant mass spectrum in HERA-I data at $M=1522\pm6$~MeV;
 $221\pm48$ events above the background were seen.
 In a similar search \cite{pos:hep2005:086} H1 did not observe any excess
 over the background.

 Since the $\Theta^+$ belongs to an hypotetical antidecuplet of pentaquarks
 with spin-$\frac{1}{2}$, two more exotic states are foreseen:
 $\Xi_{3/2}^{--}$ and $\Xi_{3/2}^+$ with quark content $ddss\bar{u}$ and
 $uuss\bar{d}$, respectively.
 NA49 at CERN SPS reported the observation of $\Xi_{3/2}^{--}$ and
 $\Xi_{3/2}^0$, whereas searches made by other experiments were negative.
 ZEUS performed such a search but did not observe any
 signal \cite{pl:b610:199}.

 The anti-charmed pentaquark $\Theta^0_c$ ($uudd\bar{c}$) has been observed
 by H1 \cite{pl:b588:17}: a peak of $50.6\pm11.2$ events was seen in the
 $D^{*-}p$ and $D^{*+}\bar{p}$ decay channels.
 However, ZEUS has not observed any excess over the
 background \cite{epj:c38:29}.

 No final conclusion can be drawn on pentaquarks.

\section{Conclusion}
 In this report a review of recent results from HERA has been presented.

 The most relevant result is probably the measurement of the proton
 struction functions and the extraction of parton distribution functions.
 Almost all the results were obtained by using HERA-I data only,
 so we expect a big improvement in the precision when
 HERA-II data will be included.

% BibTeX users please use
\bibliographystyle{unsrtnat}
\bibliography{parenti_ifae06}
%
% Non-BibTeX users please follow the syntax
% the syntax of "referenc.tex" for your own citations
%\input{parenti_ifae06_ref}
%%%%%%%%%%%%%%%%%%%%%%%%%%%%%%%%%%%%%%%%%%%%%%%%%%%%%%%%%%%%%%%%%%%%%%
\printindex
\end{document}